\documentstyle[art11]{article}
\begin{document}
\title {$SO(4)$ invariant basis functions for strongly\\
correlated Fermi systems}
\vskip 3cm
\author{Mario Salerno\\
\\
salerno@csied.unisa.it
\thanks{Istituto Nazionale di Fisica per la  Materia and
Department of Theoretical Physics, University of
Salerno, 84100 Salerno, Italy.}}
\maketitle

\begin{abstract}
\noindent We show how to construct  $SO(4)$ invariant
functions for strongly correlated Fermi systems
on lattices of finite sizes.  We illustrate the
method on the case of the 1D Hubbard chain with
four and with six sites.
\end{abstract}
\newpage

During the past years a great deal of interest
has been devoted to the study of strongly
correlated Fermi systems because of their
possible role as models of high $T_c$
superconductivity\cite{and87}. 
Among these systems, the Hubbard model is certainly the
simplest non trivial model for interacting
electrons in a solid. In spite of its apparent
simplicity, the mathematical and physical
properties of this model, in dimensions higher
than one, are still poorly understood.  On the
other hand, the huge dimension of the Hilbert
space ($4^f$ for a lattice of $f$ sites)
severely restricts numerical studies to
clusters of small sizes.  In order to reduce the
dimension of this space one may use the
symmetry properties of the Hamiltonian.  Basis
functions which account for the conservation of
the number operator, for the conservation of the $z$
component of the total spin $S$ and for the
translational invariance, were constructed by many
authors \cite{white92,seg94}.
For the Hubbard model on bipartite lattices, however,
one would like to use as basis functions the simultaneous
eigenfunctions of the $SO(4)$
(spin $S$ and pseudospin $J$) algebra as well
as of the translation operator.

The aim of the present paper is to show how to construct
$SO(4)$ invariant functions for strongly correlated Fermi
systems on arbitrary lattices. As a result
we give a set of rules for constructing
such functions which are easy to implement on a
computer in terms of symbolic languages. We
illustrate the method on the example of the
Hubbard chain with $f=4$ and $f=6$ sites.  For
$f=4$ we find that the $SO(4)$ diagonalization
decomposes the original $256*256$ hamiltonian
matrix into blocks of dimension less or equal to
four. The ground state at half filling is
obtained by diagonalizing a $3\times3$ block
corresponding to $S=J=0 $ and with momentum equal
to $k=\pi$. In the case of the $f=6$ chain we
find that the ground state at half filling is
obtained by diagonalizing a $14*14$ band matrix
corresponding to the $S =0, J = 0, k=\pi$ space,
with the width of the band equal to six.
Finally, at the end of
the paper, we estimate the sizes of the $SO(4)$
blocks as a function of $f, J, S$.

\noindent Let us start by introducing the Hubbard Hamiltonian as
\begin{equation}
\label{hub}H=-t\sum_{\sigma
{i}}^f (c_{i_\sigma }^{\dagger }c_{{i+1}_\sigma }
+ c_{{i+1}_\sigma }^{\dagger }c_{{i}_\sigma })
\;{+\;}U\sum_i^fn_{i_{\uparrow }}n_{i_{\downarrow }},
\end{equation}
where $c_{i_\sigma }^{\dagger },c_{j_\sigma }$,
($\sigma =\uparrow $ or $\downarrow$ ), are usual
fermion creation and annihilation operators.  As
well known, this model is invariant under $SU(2)$
spin rotations with generators
\begin{equation}
\label{spin}
S^{+}=\sum_{i}^f c_{i_\uparrow }^{\dagger
}c_{i_\downarrow },
\quad S^{-}=(S^{+})^{\dagger}, \quad S_z= {\frac 1 2}(N_\uparrow -
N_\downarrow).
\end{equation}
and under $SU(2)$ pseudo-spin rotations with
generator
\begin{equation}
\label{pspin}
J^{+}=\sum_{i}(-1)^i c_{i_{\uparrow}}^{+}
c_{i_{\downarrow}}^{+},
\quad\quad
J^{-}=\sum_{i}(-1)^i c_{i_{\downarrow}}
c_{i_{\uparrow}} \quad\quad J_z= {\frac{N-f}2}.
\end{equation}
One readily checks that $S, S_z, J, J_z, H$ form
a complete set of compatible observables. Moreover, from the
expression of $S_z, J_z$ in Eq.s( {\ref{spin},\ref{pspin}})
it is evident that the states of the system must have spin and
pseudospin both integer or both half integer, this giving a
$Z_2$ reduction which leads to a global $SO(4)$
symmetry \cite{YZ}.  To construct the basis
functions which span the irreducible
representations of $SO(4)$ we will use the
representation theory of the permutation group.
As well known, Young tableaux are natural objects
for this purpose. To this end we remark that the commutation
between the permutation operations and the spin
rotations implies that the irreps of $S_f$
are automatically irreps of the  $SU(2)$ spin algebra.
This was used in ref.\cite{zp96,pms96}) to solve
the Hubbard model with unconstrained hopping for
clusters of arbitrary sizes.  Here we extend this
result to the $SO(4)$ symmetry and to arbitrary
lattices.  To this end let us denote with the symbols $|3>,
|2>,|1>,|0>$ the four states on a given site
(respectively the doubly occupied, the single
occupied spin up and spin down states and the
vacuum) and let us introduce the quantum number
\begin{equation}
M = {\frac 3 2} (f-2 J_z) + S_z .
\label{M}
\end{equation}
Eigenmanifolds of $S^2, S_z, J^2, J_z$ can be
constructed from the irreps of $S_f$ by
considering all possible partitions
$(m_1,m_2,...,m_f)$ of $M$ into $f$ parts
(compatible with $N, S_z, J_z$), with
$m_i=0,1,2,3$, $m_1 \geq m_2 \geq ...\geq
m_f$, and by filling the quanta $m_i$ of each
partition in the boxes of a Young tableaux
according to the following rules.

\noindent i) The quanta must not increase
when moving from left to right in a row or when
moving down in a column.

\noindent ii) The quanta referring to spin
up and spin down states ($m=1, 2$) must not
appear more than once in a row.

\noindent iii) The quanta referring to doubly
occupied states or to empty states ($m=3, 0$)
must not appear more than once in a column.
\noindent

To pass from Young tableaux to states one usually
apply standard Young symmetrizer and
antisymmetrizer operators \cite{ha62}. Here,
however, due to the fermion realization of the
spin algebra, one must include minus signs in the
permutations which involve the interchange of two
fermions to account for the Pauli exclusion
principle.  Moreover, an additional $(-1)^j$ sign
is required for the permutations which involve
doubly occupied states at site $j$ to balance
the alternating sign of the pseudospin algebra
(without this sign  the states would
be eigenstates of $S, S_z, J_z$, but
not necessarily of $J$). To select among the
filled Young tableaux the ones corresponding to
highest weight vectors of $SO(4)$, (actually highest
weights for spin and lowest weights for
pseudospin) we use the following criterion.  Let
us call  the change of a spin down into a spin
up a $1-2$ flip and the change of a double
occupied state into an empty state a $3-0$ shift.
We note that $S^+$ just performs $1-2$ flips
while $J^-$ performs $3-0$ shifts.  It is clear
that the filled tableaux which become
unconsistent with the filling rules after 1-2
flips and 3-0 shifts are annihilated by both
$S^+$ and $J^-$ i.e. they are highest-lowest
weight vectors of the SO(4) algebra.
On filled tableaux then, we perform
1-2 flips (i.e. turn a 1 into a 2 in all possible manners)
to select the highest weights vectors of the spin
$SU(2)$, and 3-0 shifts to select the lowest weights of
the pseudospin algebra.  Note that
according to our rules, double occupied states
must be in the first row of a tableaux in
consecutive order from the first box on the
left. One can shift out a double occupied state of the first
row without changing the spin symmetries, by
moving the quanta one unit to the left and inserting
a zero in the end box of the row, or by moving the quanta in the first
column one unit upward and inserting a zero in
the end box of the column.  The tableaux that
become unconsistent with the filling rules under
both $1-2$ flips and $3-0$ shiftings are the highest-lowest
weight vectors of the $SO(4)$ algebra. We remark
however, that by  $1-2$ flips and  $3-0$ shifts two
different tableaux can be related to the
same tableau of higher (lower) spin (pseudospin).
In this case, linear combinations of the
corresponding states must be taken.  Having
constructed the  $SO(4)$ invariant states we  come to
the problem of projecting them on a particular
subgroup, say $G$, of $S_f$
(we recall that any discrete group is a subgroup
of $S_f$). To this end we denote by $ D(R),R\in
S_f$ the irreps of $S_f$. A representation of $G$
is readily obtained by selecting among the
matrices $D(R)$ those corresponding to elements
of $G$. These representations however are in
general reducible i.e. they can be expressed in
terms of irreps $D^{(\nu )}$ of $ G$ as
$D(R)=\sum_\nu c_\nu D^{(\nu )}(R)$ with $c_\nu $
non negative integers counting the number of
times $D^{(\nu )}$ appears in $D$.  By denoting
with $g_i$ the number of elements in the class
$K_i$ of $G$ and with $g$ the order of this
group, one easily express the integers $ c_\nu $
in terms of the characters $\chi ,\chi ^{(\nu )}$
of respectively $ S_f$ and $G$ as
\begin{equation}
\label{splitt}c_\nu =\frac 1g\sum_ig_i\chi _i^{(\nu )\,*}\chi _i.
\end{equation}
This gives the splitting of the irreps of $S_f$
into the irreps of $G$ (here $*$ denotes complex
conjugation).  The eigenfunctions $\psi$ of $S^2,
J^2$, corresponding to the above highest-lowest weight
states, are then projected on the $\nu
-$th irrep of $ G$ by using the projection
operator $P^{(\nu )}$
\begin{equation}
\label{project}\psi ^{(\nu )}=
P^{(\nu )}\psi \equiv \frac{n_\nu }g\sum_R\chi
^{(\nu )\,*}(R)\,\;U_R\cdot \psi
\end{equation}
where the sum is over all the elements $R$ of the
$S_f$ subgroup ($n_\nu $ is the dimension of the
$\nu$-th irrep of $G$, $\chi^{(\nu )}(R)$ the
corresponding characters and $U_{R}$ the operator
associated to the group element $R$) \cite{ha62}.
By taking $G$ to be the subgroup corresponding to
the lattice translations $T_n$  we get the simultaneous
eigenfunctions of $S^2, J^2, T_n$ with respect to
which the Hamiltonian acquires a block diagonal
form.  To illustrate the method let us consider
the case of the Hubbard model on a 1D periodic
chain with $f=4$.  This case corresponds to take
the cyclic subgroup $C_4,$ of $S_4$ which is an
abelian group with 1D irreps denoted by $A, B,
E_1, E_2$.  The irreps $E_1,E_2$ are one the
complex conjugate of the other thus they are
associated to the same energy level. We can take
advantage of this accidental
degeneracy by considering $E_1,E_2$
equivalent to a single irrep $E$ of dimension two
(note that this degeneracy is connected with the
time reversal invariance of the Hubbard
hamiltonian).  In Table 1 we have reported the
filled tableaux associated to the highest-lowest
weight vectors of the spin and pseudo-spin
algebra constructed with our rules, for even
values of the fillings (the odds fillings  follow
similarly and will be omitted for brevity).
In this table we have also reported
the splitting of the tableaux under the
translation group. Note that, due to  the presence of the
$(-1)^j$ sign in the Young operators, the even and odd
representations of $C_4$ are interchanged  in the tableaux
containing odd numbers of double occupied states.

From Table 1 we see that the N=0 sector
contains only one state (the vacuum) with eigenvalue
$E_0=0$. This state is a singlet under $S$ and a
quintet under $J$. The application of the pseudospin
rising operator to it, generates an eigenstate of H
of the $N=2$ sector with $S=0$, $J=2$ but with
$J_z=1$. The corresponding  eigenvalue is  $E_2=U$
(the state does not appear in our table since it is not
a lowest weight). By further
applications of $J^{+}$ to the vacuum on gets all its
descendents to higher
(even) fillings
\begin{equation}
E_N= E_0+{N\over 2} U ,\quad\quad N=2,4, ..., 2f.
\label{gs}
\end{equation}
This is a general feature of the $SO(4)$
diagonalization i.e. with the rising operator
$J^+$ we can map all the spectrum computed at
lower fillings $N$ to the higher fillings $N+2 n,
\quad n=1,2,...f-N/2$.
From Table 1 we also see that the $N=2, S=1$ sector
decomposes into two $1*1$ blocks ( A, B states)
and one $2*2$ block (E states) for a total of 18
states, while the $N=2, S=0$ sector decomposes
into three blocks, two are $2*2$ (B, E states)
and one is $3*3$ (A states) for a total of $9$
states. The dimension of the $N=2$ sector is then
$18+9+1=28$ (note the addition of the state $J=2,J_z=1$, coming
from the $N=0$ sector).  Similarly we see
that the $N=4$ sector contains $10$ states with $S=0$,
9 states with $S=1$ and one state with $S=2$, for
a total of 42 states. If one adds these states to
the ones coming from lower (even) fillings one gets
the total dimension of the $N=4$ sector as
$42+18+9+1=70$. These numbers just coincide with
the ones obtained from the usual formula
$d_N=\left(_{N}^{2 f}\right)$ for  N electrons on 2f sites.
By rising these states to higher fillings we
get the total dimension of the
Hilbert subspace corresponding to the even fillings
as $1*5+27*3+42 = 128$. One checks that this
number is just the sum of all the tableaux in
Table 1 taken with their multiplicities (i.e. the
product of the tableaux dimension times the spin
and pseudospin degeneracies). This checks the
completeness of our basis (the same analysis performed
on odd fillings gives others $128$ eigenvalues for a
total of  $4^4$ states).  In Tables 2a, 2b we have
reported the $SO(4)$ block
decomposition of the hamiltonian matrix
corresponding to the states listed in Table 1.
Note that the eigenvalues at higher (even) fillings are obtained
by diagonalizing the same blocks of Tables 2 but  with an $U$
added to the diagonal elements for each
application of $J^{+}$ (the blocks at
half filling obviously do not have descendents since $J=0$).
As expected, the ground state
at half filling is a singlet with $S=J=0$ \cite{lmat}.
It is of interest to note that this state has momentum $k=\pi$
(B state) and comes from the splitting
of the  same tableaux which characterizes the ground state of the $S_f$
invariant Hubbard model (see \cite{zp96,pms96}).
One can easily compare these results with direct
numerical diagonalizations in the full Hilbert space.
Thus, for example, for $N=4$ ($d_4=70$) one finds
that at $t=.75,U=1.5$ the ground state
is a singlet with energy $E=-2.12132$.
This is just the same value  obtained from the $S=J=0, B$ block
of Table 2a.

Using the $SO(4)$ symmetry we have also
investigated a chain with $f=6$ sites. Here for
brevity we report only on the ground state at
half filling (more details will be given elsewhere).
In this case we find that the ground state is
obtained by diagonalizing a $14 \times 14$ band matrix
corresponding to the $S=J=0, k=\pi$ block reported in Table 3.
In this table the diagonal elements of the block
are at the bottom while the elements above the
diagonal (moving along columns), in the rows
above. It is remarkable that the width of the band is just
equal to the filling (this feature is true also for other
fillings  $N<6$ \cite{msfai96}).
One can check that the eigenvalues of the
$SO(4)$ block in Table 3 are just the same as those 
obtained by diagonalizing the $400*400$ 
matrix of the  $S_z=0$ space 
(for example, the ground state energy 
at $t=1, U=1$ in both cases is $E=-6.60116$).

In closing this paper we shall estimate
how the size of the $SO(4)$ blocks will grow with $f$.
This can be done by counting the number $d_{f,S,J}$
of highest-lowest vectors of $SO(4)$ for a fixed $J,S$. We find that
\begin{equation}
d_{f,S,J}=\left( _{c_{+}}^f\right)
\left[ \left( _{c_{-}}^f\right) +\left(
_{c_{-}-2}^f\right) \right] -\left(
_{c_{-}-1}^f\right) \left[ \left(
_{c_{+}+1}^f\right) +\left( _{c_{-}-1}^f\right)
\right].
\label{dso4}
\end{equation}
where $c_{\pm} = {\frac f2} \pm (S \mp J)$ (the
derivation of this formula will be given
elsewhere \cite{msfai96}).  One readily sees that Eq.(\ref{dso4})
reproduces the correct number of states
associated to the filled tableaux reported in
Table 1. One also checks that these states, taken
with their multiplicity, reproduce the
full Hilbert space i.e.
\begin{equation}
\sum_{S=0}^{f/2}\sum_{J=0}^{f/2}S(S+1)J(J+1)d_{f,S,J}=4^f.
\label{comp}
\end{equation}
By assuming an equal splitting of the $S_f$
representations into the irreps of the
translation group, we estimate the dimension
of the blocks as $d_{f,S,J}/f$.  Thus, for a
chain with $10$ sites the block
characterizing the ground state at half filling
is $560\times 560$ while for a chain of
$20$ sites is $5*10^7\times 5*10^7$. This shows
that the $SO(4)$ block diagonalization allows to
reduce of about two order of magnitude the sizes
of the matrices constructed by using just the $S_z$
and the translational symmetry (this last being
${\frac 1 f} \left(_{N_{\uparrow}}^f\right)
\left(_{N_{\downarrow}}^f\right)$).
Moreover, it is likely that (at least in the 1D case) the
band structure observed at $f=6$ exists also at higher
values of $f$.
The possibility that the $SO(4)$ diagonalization
leads to band matrices for arbitrary values
of $f$ and $N$ is presently under investigation.
\vskip .5cm
\noindent{\bf{Acknowledgments}}

\noindent I wish to thank
Prof. M. Rasetti and Prof. A.C.Scott for encouragements.
Financial support from the INFM and from the
INTAS grant n.o 93-1324 is acknowledged.
\newpage

\noindent{\bf{Table Captions}}
\vskip 1cm
\noindent Table 1

\noindent Decomposition of the filled Young tableaux
corresponding to highest-lowest weight vectors of
$SO(4)$ for $f=4$ and $N=0,2,4$. The
splitting in terms of the irreps A, B, E, of the
group $C_4$ are shown.
\vskip .5cm
\noindent Table 2a

\noindent $SO(4)$ decomposition of the Hamiltonian matrix
corresponding to the highest weight vectors
of the $N=2$ states in Table 1.

\vskip .5 cm
\noindent Table 2b

\noindent $SO(4)$ decomposition of the Hamiltonian matrix
corresponding to the highest weight vectors
of the $N=4$ states in Table 1.

\vskip .5 cm
\noindent Table 3

\noindent Band structure of the $SO(4)$ block ($J=S=0,k=\pi$)
characterizing the ground state of the Hubbard
chain with six sites. The diagonal elements of the block are reported in
the row at the bottom while the elements above the
diagonal (moving along columns) in the rows
above.


\vskip 5cm
\begin{figure}
\begin{picture}(450,450)
\put(-10,550){\line(550,0){435}}
\put(-10,445){\line(550,0){435}}
\put(-10,315){\line(550,0){435}}
\put(-10,150){\line(550,0){435}}
\put(30,150){\line(0,800){430}}
\put(110,150){\line(0,800){430}}
\put(265,150){\line(0,800){430}}
\put(425,150){\line(0,800){430}}
\put(60,573){$N=0$}
\put(170,573){$N=2$}\put(310,573){$N=4$}
\put(60,560){$J=2$}
\put(170,560){$J=1$}\put(310,560){$J=0$}
\put(-5,490){$S=2$}\put(-5,375){$S=1$}
\put(-5,230){$S=0$}
\put(355,183){\framebox(13,13){2}}
\put(368,183){\framebox(13,13){1}}
\put(355,170){\framebox(13,13){2}}
\put(368,170){\framebox(13,13){1}}
\put(388,175){$A, B$}
\put(290,183){\framebox(13,13){3}}
\put(303,183){\framebox(13,13){3}}
\put(290,170){\framebox(13,13){0}}
\put(303,170){\framebox(13,13){0}}
\put(323,175){$A, B$}
\put(320,225){\framebox(13,13){3}}
\put(333,225){\framebox(13,13){2}}
\put(348,225){\framebox(13,13){1}}
\put(320,212){\framebox(13,13){0}}
\put(368,220){$A, E$}
\put(325,281){\framebox(13,13){3}}
\put(338,281){\framebox(13,13){1}}
\put(325,268){\framebox(13,13){2}}
\put(325,255){\framebox(13,13){0}}
\put(358,280){$B, E$}
\put(290,351){\framebox(13,13){3}}
\put(303,351){\framebox(13,13){2}}
\put(290,338){\framebox(13,13){2}}
\put(290,325){\framebox(13,13){0}}
\put(319,340){$B, E$}
\put(360,351){\framebox(13,13){2}}
\put(373,351){\framebox(13,13){1}}
\put(360,338){\framebox(13,13){2}}
\put(360,325){\framebox(13,13){2}}
\put(395,340){$A, E$}
\put(360,398){\framebox(13,13){3}}
\put(373,398){\framebox(13,13){2}}
\put(360,385){\framebox(13,13){2}}
\put(373,385){\framebox(13,13){0}}
\put(395,395){$A,B$}
\put(290,411){\framebox(13,13){3}}
\put(290,398){\framebox(13,13){2}}
\put(290,385){\framebox(13,13){2}}
\put(290,372){\framebox(13,13){0}}
\put(315,395){$A$}
\put(320,506){\framebox(13,13){2}}
\put(320,493){\framebox(13,13){2}}
\put(320,480){\framebox(13,13){2}}
\put(320,467){\framebox(13,13){2}}
\put(345,495){$B$}
\put(120,188){\framebox(13,13){2}}
\put(133,188){\framebox(13,13){1}}
\put(146,188){\framebox(13,13){0}}
\put(120,175){\framebox(13,13){0}}
\put(165,185){$B,E$}
\put(195,188){\framebox(13,13){3}}
\put(208,188){\framebox(13,13){0}}
\put(221,188){\framebox(13,13){0}}
\put(195,175){\framebox(13,13){0}}
\put(238,183){$A,E$}
\put(150,225){\framebox(13,13){2}}
\put(163,225){\framebox(13,13){1}}
\put(176,225){\framebox(13,13){0}}
\put(189,225){\framebox(13,13){0}}
\put(210,228){$A$}
\put(168,270){\framebox(13,13){2}}
\put(181,270){\framebox(13,13){1}}
\put(168,257){\framebox(13,13){0}}
\put(181,257){\framebox(13,13){0}}
\put(205,269){$A\,,\,B$}
\put(40,225){\framebox(13,13){0}}
\put(53,225){\framebox(13,13){0}}
\put(66,225){\framebox(13,13){0}}
\put(79,225){\framebox(13,13){0}}
\put(98,230){$A$}
\put(145,400){\framebox(13,13){2}}
\put(158,400){\framebox(13,13){0}}
\put(171,400){\framebox(13,13){0}}
\put(145,387){\framebox(13,13){2}}
\put(190,400){$B\,,\,E$}
\put(145,356){\framebox(13,13){2}}
\put(158,356){\framebox(13,13){0}}
\put(145,343){\framebox(13,13){2}}
\put(145,330){\framebox(13,13){0}}
\put(190,356){$A\,,\,E$}
\put(180,-20){Table\, 1}
\end{picture}
\end{figure}
\newpage
Table 2a

\begin{picture}(300,300)
\put(0,65){\line(300,0){300}}
\put(40,-80){\line(0,200){175}}
\begin{tabular}{rrrrrrr}
$S=0$ & & $N=2, J=1$ & & $N=4, J=0$ \\
\\A & & $ \left(
\begin{array}{ccc}
-\frac 83t & \frac 23\sqrt{2}t & \frac 43\sqrt{3}t \\
\ast  & \frac 83t & \sqrt{\frac 83}t \\
\ast  & * & U
\end{array}
\right)
$ & & $ \left(
\begin{array}{ccc}
U & 2\sqrt{3}t & 2t \\
\ast  & 2U & 0 \\
\ast  & * & 0
\end{array}
\right) $ \\ \\ B & &$
\begin{array}{c}
E=0 \\ E=0 \end{array}$ & & $ \left(
\begin{array}{ccc}
2U & 0 & -2t \\
\ast  & 0 & -2\sqrt{3}t \\
\ast  & * & U
\end{array}
\right)
$
\\ \\
E & &$E_{\pm }=\frac U2\pm \sqrt{4\,t^2+\left( \frac
U2\right) ^2}$ & & $
\begin{array}{c}
E=U-2t \\ E=U+2t
\end{array}
$
\end{tabular}
\end{picture}
\newpage
Table 2b

\begin{picture}(300,300)
\put(5,60){\line(300,0){240}}
\put(40,-70){\line(0,200){150}}
\[
\begin{tabular}{rrrrrrr}
S=1 & & N=2, J=1 & & N=4, J=0
\\ \\
A & & $E=0$ & &
$\left(
\begin{array}{ccc}
0 & \sqrt{\frac 83}t & \frac 43 \sqrt{3} t \\
\ast  & U-\frac 83 & -\frac 23 \sqrt{2} t\\
\ast  & * & U + \frac 83 t
\end{array}
\right)$
\\ \\
B & & $E=0$ & & $
\begin{array}{c}
E = U\\ E = U
\end{array}
$
\\ \\
E & & $E_{\pm }=\pm 2t$ & & E$_{\pm }=\frac U2\pm 
\sqrt{4 t^2+\left( \frac U2\right) ^2}$
\end{tabular}
\end{picture}
\newpage
TABLE 3
\vskip 2cm
$$
\begin{array}{cccccccccccccc}
& & & & & 0 & 0 & 3\sqrt{2}t & 3t & \frac4{\sqrt{3}}t &
\frac 3{\sqrt{2}}t & 0 &
\sqrt{10}t & 0 \\
& & & & 0 & 0 & 0 & 0 & 0 & \frac 1{\sqrt{2}}t &
\sqrt{\frac 32}t & 0 & 0 & 0 \\
& & & 0 & \sqrt{3}t & 0 & 0 & \sqrt{\frac
23}t & 0 & \frac 1{\sqrt{6}}t & 0 & \sqrt{2}t &
\sqrt{5}t & \frac 4{\sqrt{5}}t \\
& & 4t & 0 & 0 & 2\sqrt{2}t & 0 & t & 0 & 0 & t & 0 &
\frac 3{\sqrt{5}}t & 0 \\ & 0 & t & 2t & 0 &
0 & 0 & \frac{\sqrt{3}}3t & 2\sqrt{2}t & -t & 0 &
0 & 0 & 0 \\ 0 & 0 & U & U & U & U & U & 2U & 2U
& 2U & 2U & 2U & 3U & 3U
\end{array}
$$
\end{document}